\documentstyle[twocolumn,prl,aps,epsf]{revtex}

\begin{document}

\draft

\title{Critical Temperature $T_{\rm c}$ versus Charging Energy $E_{\rm c}$
in Molecular-Intercalated Fullerenes
}

\author{
Chikao Kawabata,$^{1}$
Nobuhiko Hayashi,$^{2}$
and
Fumihisa Ono$^{3}$
}

\address{$^{1}$Division of Liberal Arts and Sciences,
Okayama University, Okayama 700-8530, Japan\\
$^{2}$Computer Center,
Okayama University, Okayama 700-8530, Japan\\
$^{3}$Department of Physics,
Okayama University, Okayama 700-8530, Japan
}

\date{\today}

\maketitle

\begin{abstract}
   We study the recently discovered 117-Kelvin
superconducting system C$_{60}/$CHBr$_3$ of the field-effect transistor
and propose a possibility that
the intercalation molecule CHBr$_3$ plays a role of
an electric capacitor
in the C$_{60}$ fullerene superconductor,
which contrasts rather with an expectation
that the intercalation molecule in that system
acts as simple spacer molecule.
   Estimating the critical temperatures $T_{\rm c}$ for several
C$_{60}$/X (X: intercalation molecule),
we suggest searching for intercalation molecules
with large molecular polarizabilities,
in order to attain higher $T_{\rm c}$
in the synthesis of fullerene superconductors
and to more systematically develop high $T_{\rm c}$
superconducting electronic devices.
\end{abstract}

\pacs{PACS numbers: 74.62.Bf, 74.70.Wz, 74.62.-c, 74.10.+v}

   Much attention has been focused on
C$_{60}$ fullerene superconductors \cite{gunnarsson97}.
   Quite recently,
Sch\"on {\it et al.} \cite{schoen01a} found a
hole-doped C$_{60}$ superconducting system C$_{60}/$CHBr$_3$,
which exhibited very high critical temperature
$T_{\rm c}=117$ K
at ambient pressure.
   An exciting race toward room-temperature superconductivity
is now about to begin \cite{dagotto01} and would heat up.
   The superconducting system reported by Sch\"on {\it et al.}
is  a C$_{60}$ fullerene superconductor intercalated with CHBr$_3$ molecules,
in which holes can be induced by gate doping
in a field-effect transistor geometry \cite{schoen01a,schoen00a}.
   The organic field-effect transistor systems
such as this fullerene superconductor
are expected to open up a way to tunable superconducting devices
as switches, light emissions, Josephson junction circuits,
quantum computing circuits, etc.\ \cite{devices}.
   Searching for superconducting field-effect transistor systems
with higher $T_{\rm c}$
is important in the viewpoint of such application.

   The neutral molecule CHBr$_3$ in that molecular-intercalated
fullerene superconductor
acts as spacer molecule which expands the separation between C$_{60}$
molecules.
   Indeed the critical temperatures of
the C$_{60}$ fullerene superconductors
increase with the lattice constants $a$ of the C$_{60}$
crystals \cite{schoen01a,schoen00a};
C$_{60}$ without intercalation
($a=14.16$ \AA, $T_{\rm c}=52$ K),
C$_{60}/$CHCl$_3$ ($a=14.28$ \AA, $T_{\rm c} \simeq 81$ K), and
C$_{60}/$CHBr$_3$ ($a=14.43$ \AA,
$T_{\rm c}=117$ K).
   Then, one of the promising courses of attaining higher $T_{\rm c}$
is certainly to search for new spacer
molecules which expand the lattice constant
larger \cite{schoen01a,schoen00a}.

   From a different point of view, however,
we will point out another possibility that
the intercalation molecules play a role of electric capacitors
in the C$_{60}$ fullerene superconductors.
   By estimating $T_{\rm c}$ for
several C$_{60}$/X  (X: intercalation molecule),
we suggest searching for intercalation molecules
with large molecular polarizabilities
in order to attain higher $T_{\rm c}$
in the future synthesis of fullerene superconductors.
   It might become a motivation for searching and synthesizing
new intercalation molecules
with larger molecular polarizability
in the field of chemistry.

   We model those C$_{60}$ fullerene superconductors
as quantum-capacitive
Josephson junction arrays (JJA)
with coupled superconducting grains
on a lattice,
following phenomenological Kawabata-Shenoy-Bishop (KSB)
theory \cite{kawabata94a,kawabata95a,kawabata95b},
originally proposed to high $T_{\rm c}$ cuprates.
   In the present JJA model,
the superconducting grains on a lattice
are composed of C$_{60}$ molecules.
   When the Cooper pairs tunnel
between the grains, inter-grain charge unbalance is induced.
   The charge unbalance gives rise to the electric field
between the grains, and it costs the electrostatic energy
and destroys the phase coherence of
superconductivity \cite{kawabata94a,kawabata95a,kawabata95b,simanek}.
   This energy is called as the charging energy $E_{\rm c}$ and it
depresses $T_{\rm c}$.
   On the other hand,
the Josephson coupling energy $E_{\rm J}$ of the system
corresponds to a ``bare'' $T_{\rm c}$ without any influence of
the charging energy.
   The bare $T_{\rm c}$ (or $E_{\rm J}$) may be determined microscopically
by
the Cooper pairing mechanism such as
the electron-phonon couplings \cite{gunnarsson97}
and the kinetic energy lowering due to
the local hole concentration \cite{hirsch00}.
   In the present model,
experimentally observed differences in $T_{\rm c}$ between
hole-doped and electron-doped
C$_{60}$ fullerene superconductors intercalated
with the same molecule \cite{schoen01a,schoen00a}
should reflect a difference in such $E_{\rm J}$.
   In any case, the present phenomenological model does not depend on
any specific microscopic mechanisms of superconductivity
which determine $E_{\rm J}$.
   The optimal $T_{\rm c}$ (or optimal $E_{\rm J}$) of each system
can be experimentally attained
with adjusting the carrier doping
owing to the field-effect transistor
technique \cite{schoen01a,schoen00a}.
   Then, we consider the charging energy $E_{\rm c}$
and discuss how high $T_{\rm c}$ can be realized in various C$_{60}$/X systems
under the condition that the optimal $T_{\rm c}$ are attained
in each system.
   An important point for our analysis is that
the intercalation molecules X in C$_{60}$/X superconducting systems
play a role of capacitors,
reducing the charging energy $E_{\rm c}$
by the electronic polarization of the molecules.

   By the electronic polarization of the neutral molecules,
the intercalation molecules such as CHBr$_3$
can reduce the charging energy $E_{\rm c}$ between
C$_{60}$ molecules.
   The smaller the charging energy $E_{\rm c}$ in JJA,
the higher the critical temperature $T_{\rm c}$,
as discussed for
high $T_{\rm c}$ cuprates \cite{kawabata94a,kawabata95a,kawabata95b}
and a diboride superconductor MgB$_2$ \cite{kawabata01,kawabata01-2}
on the basis of the same model.
   According to
experimental data \cite{duijnen98},
the molecular polarizability $\alpha$ of CHBr$_3$ (79.9 au) is indeed
larger than that of CHCl$_3$ (57.56 au),
which is in accord with the present scenario that
the intercalation molecules act as capacitors
with the capacitance $C$ ($\propto \alpha$, roughly)
which reduce
$E_{\rm c}$ ($\propto C^{-1}$)
and increase $T_{\rm c}$.

   We estimate $T_{\rm c}$ as a function of
$E_{\rm c}$.
   According to the KSB theory \cite{kawabata94a,kawabata95a,kawabata95b},
$T_{\rm c}$ is expressed, in the present JJA model, as
$E_{\rm J} / k_{\rm B}T_{\rm c}
=0.454f(\gamma_0^{-2},E_{\rm c}/E_{\rm J})$
with an unknown function $f$.
   It must reduce to
a known equation for zero charging energy ($E_{\rm c}=0$),
$E_{\rm J} / k_{\rm B}T_{\rm c}
=0.454f(\gamma_0^{-2},0)$,
which is the dimensionless critical coupling
in terms of the anisotropy parameter $\gamma_0^{-2}$
in the classical anisotropic 3D XY/JJA system \cite{shenoy94,shenoy95}
(in 3D limit $\gamma_0^{-2} \rightarrow 1$,
while in 2D limit $\gamma_0^{-2} \rightarrow 0$).
   Expanding $f$ with respect to $E_{\rm c}/E_{\rm J}$, we obtain
an approximated expression,
$k_{\rm B}T_{\rm c}
\approx 2.2 E_{\rm J}[f_0+f'_0 E_{\rm c}/E_{\rm J}]^{-1}
\approx 2.2 (E_{\rm J}/f_0)[1- (f'_0/f_0)
(E_{\rm c}/E_{\rm J})]$ \cite{kawabata94a,kawabata95a,kawabata95b},
where $f_0=f(\gamma_0^{-2},0)
\sim 1$ \cite{shenoy94,shenoy95}.
   Here, we express it as
\begin{equation}
T_{\rm c} = A (1- B E_{\rm c})
\label{eq:Tc-KSB-4}
\end{equation}
or, assuming a BCS like expression (see a discussion below),
\begin{equation}
T_{\rm c} = {\bar A} \exp \Bigl[\frac{-2}{1- {\bar B} E_{\rm c}}\Bigr],
\label{eq:Tc-KSB-5}
\end{equation}
with the phenomenological parameters
$A$, $B$, ${\bar A}$, and ${\bar B}$,
which should be determined empirically.

   The charging energy $E_{\rm c}$ is estimated
as follows.
   Considering two superconducting grains of C$_{60}$
and an intercalation molecule situated between them,
   we estimate the electric field $E$ induced when
a Cooper pair tunnels between these grains.
   The electric flux density $D$ between the grains equals to
the electric charge of the pair $2e$ over the cross section of the grain
$S$ (of the order of the square of the diameter of a C$_{60}$ molecule or
the coherence length),
namely $2e/S \equiv q = D = \varepsilon_0 E + P$,
where $\varepsilon_0$ is the vacuum dielectric constant and
$P$ the polarization due to the intercalation molecule.
   Assuming one intercalation molecule exists between the grains
and then neglecting effects of
the depolarization field, we obtain a relation
$P \approx n \alpha E/a^3$,
where $\alpha$ is the molecular polarizability of the intercalation molecule
and $n$ the number of the intercalation molecule per unit cell
which depends on the crystal structure and on the lattice constant $a$.
   Thus $E=q/ (\varepsilon_0 +\varepsilon)$,
where $\varepsilon \equiv n \alpha/a^3$.
   We obtain the capacitance $C$ between the grains as
$C=2e/Ed =(\varepsilon_0 +\varepsilon)S/d$,
where $d$ is the separation between the grains ($d \propto a$).
   Therefore, the charging energy is estimated as
$E_{\rm c}=(2e)^2/2C = (2e)^2 d/2S(\varepsilon_0 +\varepsilon)
\propto a/(\varepsilon_0 +\varepsilon)$.
  We rewrite it further as
$E_{\rm c} \sim a/(\varepsilon_0 + \varepsilon)
= a (\varepsilon_0 + n\alpha/a^3 )^{-1}
= a \varepsilon_0 (1+4\pi n\alpha_{\rm CGS}/a^3 )^{-1}$,
and finally
\begin{equation}
E_{\rm c} = \frac{\tilde a}{1+ 4\pi n\alpha_{\rm CGS}a^{-3}}
\label{eq:Ec}
\end{equation}
in arbitrary units,
where ${\tilde a} \equiv a/a({\rm C}_{60})=a/(14.16\ {\rm \AA})$ and
$\alpha_{\rm CGS}$ ($\equiv \alpha/4\pi\varepsilon_0$) \cite{kittel}
corresponds to
the molecular polarizabilities listed in Table \ref{table:Ec}.

\begin{figure}
\epsfxsize=85mm
\begin{center}
\epsfbox{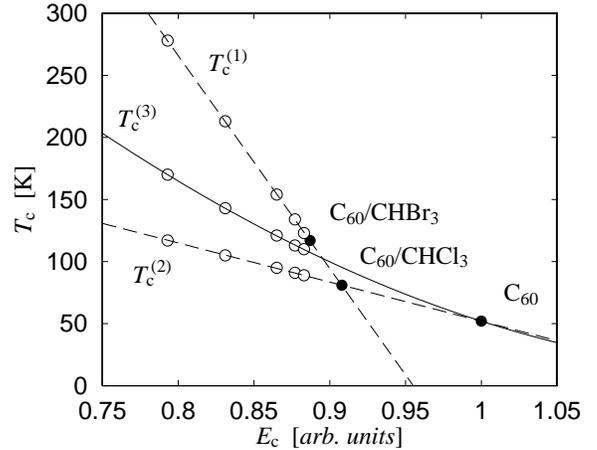}
\end{center}
\vspace{5.5mm}
%
\caption{
   Plots of critical temperature $T_{\rm c}$
versus charging energy $E_{\rm c}$.
   The critical temperatures $T_{\rm c}$ of solid circles
represent the experimental data [2,5].
   The values of $E_{\rm c}$ are
estimated on the basis of Eq.\ (\ref{eq:Ec}) (Table \ref{table:Ec}).
   Open circles on dashed lines
show results obtained for several C$_{60}$/X
by linear extrapolations
(the dashed lines $T_{\rm c}^{(1)}$ and $T_{\rm c}^{(2)}$)
based on Eq.\ (\ref{eq:Tc-KSB-4}),
and they correspond to
the results listed in Table \ref{table:Ec}.
   The solid line $T_{\rm c}^{(3)}$
and open circles on it
represent a BCS like relation
$\bigl[$Eq.\ (\ref{eq:Tc-KSB-5})$\bigr]$,
and the line is plotted such that
it goes through the solid circle for C$_{60}$
and the average point of the other two solid circles.
   Actual $T_{\rm c}$ of those C$_{60}$/X
experimentally observed
are expected to be distributed
in the region between $T_{\rm c}^{(2)}$ and $T_{\rm c}^{(1)}$.
}
\label{fig:tc}
\end{figure}

   On the basis of the above formulation,
we estimate $T_{\rm c}$ of C$_{60}$/X as follows,
for several molecules X listed in Ref.\ \cite{duijnen98}.
   The experimental values of the molecular polarizability
of CHCl$_3$ and CHBr$_3$
are $\alpha$(CHCl$_3)=57.56$ au and
$\alpha$(CHBr$_3)=79.9$ au \cite{duijnen98}.
   We obtain
$E_{\rm c}($C$_{60}/$CHCl$_3)=
{\tilde a}(1+4\pi n\alpha_{\rm CGS}/a^3 )^{-1} =0.908$ and
$E_{\rm c}($C$_{60}/$CHBr$_3)=0.887$
on the basis of Eq.\ (\ref{eq:Ec}),
where the cubic lattice with the lattice constant $a$ and $n=3$ are assumed
representatively.
   On the $T_{\rm c}$ vs.\ $E_{\rm c}$ graph,
extrapolating the straight line $\bigl[$Eq.\ (\ref{eq:Tc-KSB-4})$\bigr]$
which goes through two points
$(E_{\rm c},T_{\rm c})=(0.908,81\ {\rm K})$ for C$_{60}/$CHCl$_3$
and $(0.887,117\ {\rm K})$ for C$_{60}/$CHBr$_3$ in Fig.\ \ref{fig:tc},
namely, $T_{\rm c}=1637-1714 E_{\rm c}$,
we predict $T_{\rm c}$ of C$_{60}$/X
(we call them as $T_{\rm c}^{(1)}$)
and show the results in Table \ref{table:Ec}.
   Here, assuming an approximately constant value
$a \equiv a({\rm C}_{60}/{\rm CHBr}_3)=14.16\ {\rm \AA}$ \cite{lattice},
we estimate $E_{\rm c}$ of C$_{60}$/X
(Table \ref{table:Ec}).
   We also calculate $T_{\rm c}$
by extrapolating another straight line which goes through
another pair of points
$(E_{\rm c},T_{\rm c})=(1,52\ {\rm K})$ for
C$_{60}$ without intercalation
and $(0.908,81\ {\rm K})$ for C$_{60}/$CHCl$_3$ in Fig.\ \ref{fig:tc},
namely, $T_{\rm c}=367-315 E_{\rm c}$.
   The results are shown as $T_{\rm c}^{(2)}$ in Table \ref{table:Ec}.
   It is expected that actual $T_{\rm c}$ observed experimentally
will be distributed
in the region between $T_{\rm c}^{(2)}$ and $T_{\rm c}^{(1)}$
in Fig.\ \ref{fig:tc}.
   We also plot a solid line in Fig.\ \ref{fig:tc} representing
the BCS like relation $\bigl[$Eq.\ (\ref{eq:Tc-KSB-5})$\bigr]$
as $T_{\rm c}^{(3)}$,
which is drawn such that
the line goes through the point for C$_{60}$ without intercalation
and the average point of the other two points for
C$_{60}/$CHCl$_3$ and C$_{60}/$CHBr$_3$
$\bigl[$thus ${\bar A}=7717$ and ${\bar B}=0.6$
in Eq.\ (\ref{eq:Tc-KSB-5})$\bigr]$.
   It is noticeable that
$T_{\rm c}$ of C$_{60}$/CHI$_3$ and C$_{60}$/C$_{12}$H$_{26}$
are anticipated being especially high because of
their small $E_{\rm c}$ in Table \ref{table:Ec}, namely
large molecular polarizabilities of their intercalation molecules.

   The present JJA model which we have used is never a particular one
and it should be understood also in the context of the BCS theory.
   The charging energy $E_{\rm c}$,
which depends on the molecular polarizability
of the intercalation molecule,
 should be related to
the inter-site Coulomb repulsion $V$ in the Hubbard model.
   That is,
the intercalation molecule situated between the sites
screens the inter-site Coulomb interaction
by its electronic molecular polarization,
and then $V$ is decreased by this screening effect
due to the polarization of the molecule.
   The electron-electron repulsion,
partly determined by $V$ (or $E_{\rm c}$),
is parametrized as $\mu^{*}$
in the usual McMillan expression of $T_{\rm c}$ \cite{mcmillan68}
based on the strong-coupling BCS theory.
   We have set Eq.\ (\ref{eq:Tc-KSB-5})
as a phenomenological equation
in the form
similar to the McMillan expression of $T_{\rm c}$ \cite{mcmillan68}
by hypothesizing $\lambda \sim 1$ (strong coupling)
and $\mu^{*} \propto E_{\rm c}$.
   First principle microscopic investigations into
effects of the intercalation molecules on
$V$
in the physical quantum chemistry approach
would be interesting as future reseaches to microscopically clarify roles of
the intercalation molecules such as CHBr$_3$.

   We note that $T_{\rm c}$ of
the chemical-doping fullerides synthesized one decade ago
such as A$_3$C$_{60}$ (A=K, Rb, and Cs) \cite{gunnarsson97}
can be partly understood in terms of the present theory.
   The present theory anticipates that
the larger the polarizability of the intercalation atom,
the higher the critical temperature $T_{\rm c}$.
   The electronic polarizabilities of the ions
are 1.33, 1.98, and 3.34 \AA$^3$ for K$^+$, Rb$^+$, and Cs$^+$ respectively,
according to Ref.\ \cite{tessman53}.
   In experimental data of $T_{\rm c}$ \cite{gunnarsson97},
there seemingly exists a tendency that
$T_{\rm c}$ of Rb$_3$C$_{60}$ is larger than
that of K$_3$C$_{60}$
by comparison at the same lattice parameter.
   Moreover,
the maximum $T_{\rm c}$ of Cs$_3$C$_{60}$ is 40 K,
while $T_{\rm c}$ of K$_3$C$_{60}$ and Rb$_3$C$_{60}$
are lower than about 30 K \cite{gunnarsson97}.
   These experimental findings are indeed in accord with
the above anticipation of the present theory.
   It should be noted, however, that
for various kinds of chemical-doping fullerides \cite{gunnarsson97}
the multiple conditions (the carrier doping, lattice parameter,
crystal structure, and polarizability)
are simultaneously changed in general,
and then a universal conclusion cannot be
reached.
   In this context,
the field-effect transistor technique applied to the fullerenes
intercalated with neutral molecules
is certainly an innovation,
which enables us to
adjust freely the carrier doping
with other conditions fixed
and to investigate roles of the intercalation objects.

   As other possible role of the intercalation molecules
in the fullerene superconductors,
the contribution of
the oscillation of the intercalation molecules
to the electron-phonon coupling for the pairing
is proposed theoretically in Ref.\ \cite{bill01},
which is different from the proposition of the present theory
although both theories predict
$T_{\rm c}($C$_{60}$/CHI$_3)
>T_{\rm c}($C$_{60}$/CHBr$_3)
>T_{\rm c}($C$_{60}$/CHCl$_3)$ \cite{kawabata01-2,bill01}
$\bigl[$equivalently, in the present theory,
$E_{\rm c}($C$_{60}$/CHI$_3)
<E_{\rm c}($C$_{60}$/CHBr$_3)
<E_{\rm c}($C$_{60}$/CHCl$_3)$;
see Table \ref{table:Ec} and refer to
Eqs.\ (\ref{eq:Tc-KSB-4}) and (\ref{eq:Tc-KSB-5}) for $T_{\rm c}$
as functions of $E_{\rm c}$$\bigr]$.
   In the case of the chemical-doping fullerides intercalated
with alkali atoms,
the phonon modes due to the alkali atoms have been discussed
as a candidate for the pairing mechanism \cite{zhang91}
and have been investigated experimentally \cite{gunnarsson97}.
   The role of the intercalation molecules
proposed in Ref.\ \cite{bill01} may be
confirmed experimentally by the same experimental methods such as
the isotope effect \cite{gunnarsson97}.

   In conclusion,
we indicate that in the recently discovered superconductor
C$_{60}/$CHBr$_3$,
the intercalation molecules CHBr$_3$ can act
{\it as electric capacitors} which reduce
the charging energy $E_{\rm c}$ by the electronic polarization
among the superconducting grains of C$_{60}$ molecules.
   We predict that $T_{\rm c}$ will be increased higher
by intercalating the molecules
{\it with larger molecular polarizability}.
   An unexplored course of attaining higher $T_{\rm c}$
in fullerene superconductors
may be a search for or a synthesis of intercalation molecules
with large molecular polarizability
other than a search for intercalation molecules as spacer molecule.
   We hope this proposition will help guide design of
the superconducting field-effect transistor systems,
in order to more systematically develop high $T_{\rm c}$
superconducting electronic devices.

%
%
%
%
%
%
%



\begin{table}
\caption{
Molecular polarizabilities $\alpha_{\rm CGS}$ of
intercalation molecules X
$\bigl($experimental data listed in Ref.\ [13]$\bigr)$,
charging energies in arbitrary units
$E_{\rm c} = {\tilde a}(1+4\pi n\alpha_{\rm CGS}/a^3 )^{-1}$
$\bigl[$Eq.\ (3)$\bigr]$,
and critical temperatures
$\bigl($$T_{\rm c}^{\rm (exp)}$: experimental data [2,5],
and
$T_{\rm c}^{(1,2,3)}$: predictions of the present theory$\bigr)$
for several C$_{60}$/X.
}
\vspace{5mm}
\begin{tabular}{lp{20mm}p{20mm}p{20mm}p{20mm}p{20mm}p{20mm}}
%
    &
Molecular polarizability &
Charging energy &
Critical temperature &
Critical temperature &
Critical temperature &
Critical temperature \\
C$_{60}$/X    &
$\alpha_{\rm CGS}$ [au]  &
$E_{\rm c}$    &
$T_{\rm c}^{\rm (exp)}$ [K] &
$T_{\rm c}^{(1)}$ [K] &
$T_{\rm c}^{(2)}$ [K] &
$T_{\rm c}^{(3)}$ [K] \\
\hline
C$_{60}$                      & ---        & 1      &
52 & --- & --- & --- \\
C$_{60}$/CHCl$_3$             & 57.56    & 0.908  &
$\simeq 81$ & --- & --- & --- \\
C$_{60}$/CHBr$_3$             & 79.9     & 0.887  &
117 & --- & --- & --- \\
C$_{60}$/CHI$_3$            & 121.74   & 0.831  & --- & 213  & 105 & 143 \\
C$_{60}$/CH$_2$I$_2$        & 87.05    & 0.877  & --- & 134  & 91  & 113 \\
C$_{60}$/C$_6$H$_4$Cl$_2$   & 95.83    & 0.865  & --- & 154  & 95  & 121 \\
C$_{60}$/C$_6$H$_5$NO$_2$   & 87.19    & 0.877  & --- & 134  & 91  & 113 \\
C$_{60}$/C$_6$H$_5$Cl       & 82.67    & 0.883  & --- & 123  & 89  & 110 \\
C$_{60}$/C$_{12}$H$_{26}$   & 153.86   & 0.793  & --- & 278  & 117 & 170
%
\label{table:Ec}
\end{tabular}
\end{table}


\begin{references}

\bibitem{gunnarsson97}
O. Gunnarsson,
Rev. Mod. Phys. {\bf 69}, 575 (1997).

\bibitem{schoen01a}
J. H. Sch\"on, Ch. Kloc, and B. Batlogg,
Science {\bf 293}, 2432 (2001).

\bibitem{dagotto01}
E. Dagotto,
Science {\bf 293}, 2410 (2001).

\bibitem{devices}
J. H. Sch\"on {\it et al.},
Science {\bf 287}, 1022 (2000);
Science {\bf 288}, 656 (2000);
Science {\bf 289}, 599 (2000);
Science {\bf 290}, 963 (2000);
Science {\bf 292}, 252 (2001);
G. Wendin and V. S. Shumeiko,
Science {\bf 292}, 231 (2001).

\bibitem{schoen00a}
J. H. Sch\"on, Ch. Kloc, and B. Batlogg,
Nature {\bf 408}, 549 (2000).

\bibitem{kawabata94a}
C. Kawabata, S. R. Shenoy, and A. R. Bishop,
in {\it Bulletin of the Electrotechnical laboratory}
(AIST, Tsukuba, 1994) Vol.58, No.6, p.426.

\bibitem{kawabata95a}
C. Kawabata, S. R. Shenoy, and A. R. Bishop,
in {\it Advances in Science and Technology 8,
Superconductivity and Superconducting Materials Technologies},
edited by P. Vincenzini
(Techna Srl., Faenza, 1995) p.13.

\bibitem{kawabata95b}
C. Kawabata,
in {\it Advances in Superconductivity VII},
edited by K. Yamafuji and T. Morishita
(Springer-Verlag, Tokyo, 1995) p.233.

\bibitem{simanek}
E. ${\rm \breve{S}}$im${\rm \acute{a}}$nek,
{\it Inhomogeneous Superconductors} (Oxford Univ. Press, New York, 1994).

\bibitem{hirsch00}
J. E. Hirsch,
Physica C {\bf 341-348}, 213 (2000).

\bibitem{kawabata01}
C. Kawabata, N. Hayashi, and F. Ono,
J. Phys. Soc. Jpn. {\bf 70}, 3184 (2001).

\bibitem{kawabata01-2}
C. Kawabata, N. Hayashi, and F. Ono,
submitted to Proceedings of
the 14th International Symposium on Superconductivity (ISS 2001),
Kobe, Japan, 2001; cond-mat/0109251.

\bibitem{duijnen98}
P. Th. van Duijnen and M. Swart,
J. Phys. Chem. A {\bf 102}, 2399 (1998).

\bibitem{shenoy94}
B. Chattopadhyay and S. R. Shenoy,
Phys. Rev. Lett. {\bf72}, 400 (1994).

\bibitem{shenoy95}
S. R. Shenoy and B. Chattopadhyay,
Phys. Rev. B {\bf 51}, 9129 (1995).

\bibitem{lattice}
The expansion of the lattice constant $a$ is only 2 \%
in C$_{60}$/CHBr$_3$ as compared with $a$ in C$_{60}$ without intercalation.

\bibitem{kittel}
C. Kittel,
{\it Introduction to Solid State Physics}
(Wiley \& Sons, New York, 1996).

\bibitem{mcmillan68}
W. L. McMillan,
Phys. Rev. {\bf 167}, 331 (1968).

\bibitem{tessman53}
J. R. Tessman, A. H. Kahn, and W. Shockley,
Phys. Rev. {\bf 92}, 890 (1953).

\bibitem{bill01}
A. Bill and V. Z. Kresin,
cond-mat/0109553;
cond-mat/0110327.

\bibitem{zhang91}
F. C. Zhang, M. Ogata, and T. M. Rice,
Phys. Rev. Lett. {\bf 67}, 3452 (1991).

\end{references}
\end{document}